# A Spectrometer as Simple as a CCD Detector

Tuo Li[1, 2*], Yishi Shi[3]

[1] *School of Science, Xijing University, Xi'an, 710123, People's Republic of China,*

[2] *Engineering Technology Research Center of Controllable Neutron Source and Its Application, Xijing University, Xi'an, 710123, People's Republic of China,*

[3] *College of Opto-Electronics, University of Chinese Academy of Sciences, Beijing, 100049, People's Republic of China,*

*\*Corresponding Author: 20160046@xijing.edu.cn*

**Abstract** Spectroscopy is the most fundamental instruments in almost every field of modern science. Conventional spectrometer is based on the dispersion elements such as various gratings. An alternative way is based on the filters such as interference filters, plasmonic nanoresonators, or quantum dots. However, for any of the above two spectrometers, the high-precision grating or the absorption filter should be elaborately designed and makes it expensive. Here, we propose a third spectrometer mechanism pupil diffraction detector (PDS). Since the high-precision grating and the elaborately designed absorption filter are abandoned, the whole structure of the PDS is just as simple as a CCD detector. Thus, compared with the above two spectrometer, the structure of the spectrometer is greatly simplified and the cost of the spectrometer is sharply reduced. In addition, the PDS can ensure the spectral range and resolution simultaneously due to the reconstruction its reconstruction algorithm. A series of simulation results are shown to demonstrate the feasibility of the PDS principle. Further, the effectiveness of the PDS principle in the noisy condition is also tested. Owing to this merits small size (the volume can be controlled within 1 cm ×1 cm ×1 cm or even smaller), light weight, and low cost, we expect the inventions of PDS has great application potential such as putting on the satellite to perform space exploration, and integrates to the smartphone to realize the detection pesticide residue in the food and clinical diagnose of the disease.

## 1. Introduction

Spectrometer is one of the most significant tools in the field of modern science, such as materials, chemicals, archeology, drug testing and other fields. At present, the most common spectrometer on the market is the use of grating for spectroscopy. Although this spectrometer can get a lot of precision spectral measurement results. However, various improvements to the spectrometer have never ceased, especially in the direction of miniaturization, low cost and ease of use [1]. The absorption filter-based spectrometer is a great solution to the miniaturization, especially the quantum dot spectrometer emerged in recent years [2-11]. Quantum dot spectrometers use quantum dots as absorption filters, which is charming technique and are expected to reduce the spectrometer to coin size [12]. However, absorption filters based spectrometers (even the quantum dots spectrometer) have a nature absorption which cause light source. And the fabrication of the absorptive filter is also complicated and expensive. Here, we propose the third mechanism of spectrum detector which is as simple as a CCD - PDS.

In the PDS, conventional dispersive elements and absorption filter are abandoned. Thus, the structure of the PDS only needs to shrink the pupil of CCD detector. Because the pupil in the PDS is completely transparent, all spectral component of the spectrum can pass through the holes and then reach the detector. Thus, this method has not absorption and an ultra-wide spectral detection range. Further, the cost of the spectrometer is particularly low because the grating and absorption filter is not required any more. As the spectrometer has the advantages of low cost, light weight and small volume (the volume can be controlled within 1 cm ×1 cm ×1 cm or even smaller). This type of spectrometer will promote the development of the spectrometer and change the life of general public [13-14].

## 2. Schematic of the PDS

The schematic of PDS is shown in Figure 1. Figure 1a shows the normally structure of CCD detector which is composed of a aperture pupil and a two-dimensional light sensor array. Figure 1b shows the structure of the PDS which is composed of a hole pupil and a two-dimensional light sensor array. We can see that the only difference between the CCD detector and the PDS is that the size of the pupil is different. The structure of the PDS is almost as simple as the structure of the detector. Thus, the cost of PDS is approximately as cheap as the cost of CCD detector.

We will show how the PDS spectrometer operates shown in Figure 1c. The incident spectrum incidents perpendicularly to the hole pupil of the PDS. Due to the small size of the hole pupil, the incident spectrum is diffracted on the hole and then reaches the two dimensional sensor array. The detector records the corresponding diffraction pattern in which different spectral components are overlapped. By applying the proposed reconstructed algorithm to process the recorded diffraction pattern, the spectral can be reconstructed.

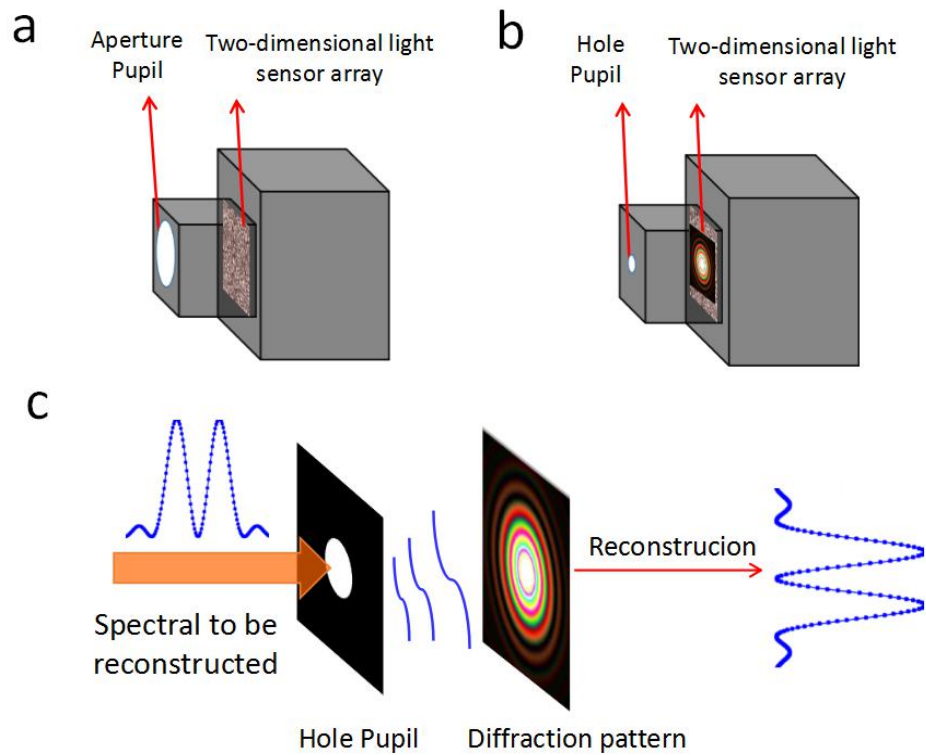

**Figure 1| a**. The structure of the CCD detector which is composed of an aperture pupil and a two dimensional light sensor array. **b**. The structure of the PDS is composed of a hole pupil and a two-dimensional light sensor array. **c**. The operation of the PDS. The diffraction patterns of each wavelength in the spectral to be detected are overlapped on the detector.

## 3. Reconstruction algorithm of the PDS

To illustrate the reconstruction algorithm mathematically, consider an arbitrary incident spectrum

$E(\lambda_i)$ (where $\lambda_i$ is the wavelength) that is transmitted through the pupil of the PDS. Each spectral component $\lambda_i$ in the incident spectral $E(\lambda_i)$ diffracts on the pupil and generates a diffraction patterns $I_{\lambda i}(m\Delta x, n\Delta y)$ on the CCD. The CCD records the total diffraction patterns $I(m\Delta x, n\Delta y)$ which is the superposition of all the $I_{\lambda i}(m\Delta x, n\Delta y)$ (where $i=1,2,...,n_q$ is the wavelength number):

$$I(m\Delta x, n\Delta y) = \sum_{\lambda_i} I_{\lambda_i}(m\Delta x, n\Delta y) = \sum_{\lambda_i} a_{mn}^{\lambda_i} E(\lambda_i), i=1,2,...,n_q \quad (1)$$

$$E(\lambda_i) = \sum_m \sum_n I_{\lambda_i}(m\Delta x, n\Delta y) \quad (2)$$

$$a_{mn}^{\lambda_i} = \frac{I_{\lambda_i}(m\Delta x, n\Delta y)}{E(\lambda_i)} \quad (3)$$

where $(m,n)$ represents the pixel matrix coordinate of the $I_{\lambda i}(m\Delta x, n\Delta y)$. $E(\lambda_i)$ is the intensity superposition of all the pixels in $I_{\lambda i}(m\Delta x, n\Delta y)$ shown in the Equation (2). $a_{mn}^{\lambda_i}$ is that the intensity of one pixel occupies the percentage of total intensity $E(\lambda_i)$.

In Equation (1), the measured total diffraction pattern $I(m\Delta x, n\Delta y)$ is known, $E(\lambda_i)$ and $a_{mn}^{\lambda_i}$ are unknown. Here, we show how to obtain $a_{mn}^{\lambda_i}$. Since the hole in the PDS is fixed, for each wavelength, the intensity distribution of diffraction pattern $I'_{\lambda i}(m\Delta x, n\Delta y)$ is predetermined according to the Fresnel law shown in Equation 4 (Here, the relation between the $I_{\lambda i}(m\Delta x, n\Delta y)$ and the $I'_{\lambda i}(m\Delta x, n\Delta y)$ is $I_{\lambda i}(m\Delta x, n\Delta y) = c_0 I'_{\lambda i}(m\Delta x, n\Delta y)$, in which $c_0$ is a constant which does not affect the intensity diffraction patterns):

$$I'_{\lambda_i}(m\Delta x, n\Delta y) = |FrT_{\lambda,z}\{P(m\Delta x_0, n\Delta y_0)\}|^2 \quad (4)$$

where $FrT\{\}$ represents the Fresnel transformation, $P(m\Delta x_0, n\Delta y_0)$ is the transmittance function of the hole. Since the $P(m\Delta x, n\Delta y)$ is known information for a each PDS, we can obtain $I'_{\lambda i}(m\Delta x, n\Delta y)$ by applying Equation (4). Because the $I'_{\lambda i}(m\Delta x, n\Delta y)$ is known, $a_{mn}^{\lambda_i}$ can be obtained by Equation (5):

$$a_{mn}^{\lambda_i} = \frac{I'_{\lambda_i}(m\Delta x, n\Delta y)}{\sum_m \sum_n I'_{\lambda_i}(m\Delta x, n\Delta y)} \quad (5)$$

Till now, only the spectrum $E(\lambda_i)$ is unknown in Equation (1). Note that once the structure of PDS is determined, the $a_{mn}^{\lambda_i}$ is fixed parameter. In the practical use, the $a_{mn}^{\lambda_i}$ is obtained by its manufacture factory and tells to the PDS user. The PDS user does not need to measure it.

There is an alternative way to obtain the fixed parameter $a_{mn}^{\lambda_i}$ of an PDS. The specific method is that the manufacture factory of the PDS use the monochromator automatic changing wavelength to illuminate the hole and records each corresponding diffraction pattern $I'_{\lambda i}(m\Delta x, n\Delta y)$. Then, calculates the $a_{mn}^{\lambda_i}$ by using the Equation (5) and tells the users.

Every pixel of total measured diffraction pattern $I(m\Delta x, n\Delta y)$ can construct an equation like Equation (1). For a given diffraction patterns $I(m\Delta x, n\Delta y)$, there are large numbers of pixels contained. Thus, these large numbers of the pixels in $I(m\Delta x, n\Delta y)$ can constructed a linear equations group. Since the $a_{mn}^{\lambda_i}$ and $I(m\Delta x, n\Delta y)$ are known, we can obtain the spectrum of $E(\lambda_i)$ by solving the equation groups. In the ideal case, we set $n_p$ equals to the $n_q$ (where $n_p$ is the number of the pixels selected in the measured diffraction patterns $I_{\lambda i}(m\Delta x, n\Delta y)$) which produces a set of linear equations with a unique solution. However, in practice, measurement errors yield inconsistencies within the set of equations. Approximate solutions can be obtained by applying the least-squares linear regression. In such a case, a given $n_p$ no longer provides an equal number of accurately reconstructed spectral data points and so $n_q < n_p$. The larger the error, the more pixels are required to accurately reconstruct each spectral data point.

## 4. Simulation Demonstrations of PDS

A series of simulation experiments are performed to demonstrate the feasibility of the PDS. Figure 2 shows an example. Figure 2a shows the image of hole pupil (1.024mm×1.024mm). The distance between the hole and the two-dimensional sensor array is 1cm shown in Figure 2b. In the simulation demonstrations, the CCD detector applied are 256×256 pixels, the pixel size of the CCD detector is 4μm× 4μm, the spectral response of CCD detector ranges from 400nm to 700nm. Figure 2c and Figure 2d show two measured diffraction patterns corresponding to two different spectral to be detected. We select 300 pixels from the diffraction patterns to construct the equations and thus the resolution of the PDS reaches 1nm. By applying the reconstructed algorithm, the spectral can be reconstructed (Solid lines in Figure 2e and Figure 2f correspond to the original spectra to be detected; Squares in Figure 2e and Figure 2f correspond to the reconstructed spectral measured by the PDS). The reconstructed spectrum and the reference spectrum are highly coincided.

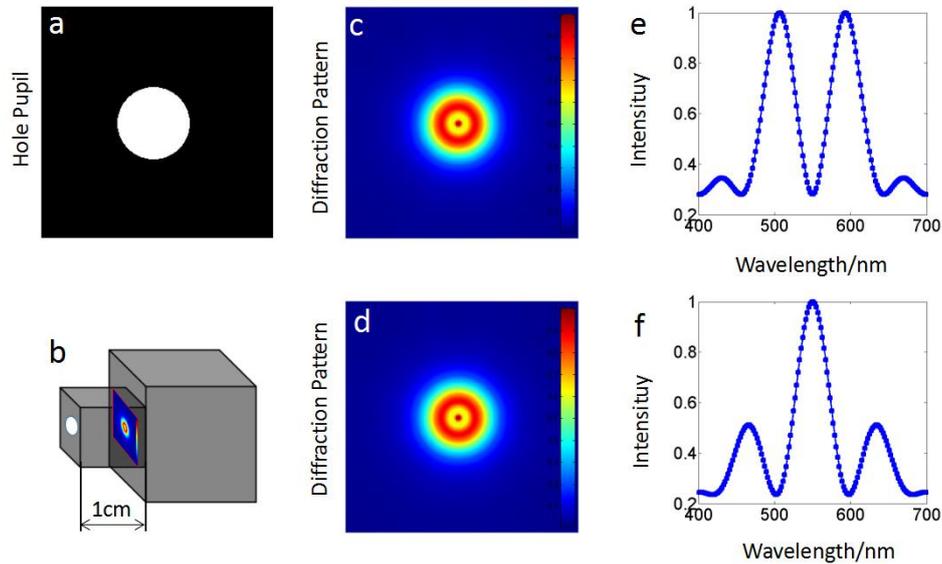

**Figure 2| Simulation experiments of the PDS**. a. the image of the hole which size is 2mm×2mm. b. The structure of the PDS in which the distance between the hole pupil and the diffraction patterns is 1cm. c and d are two diffraction patterns corresponding to two spectral to be detected. e and f are the spectrum detected by the PDS(Squares correspond to the reconstructed spectra measured by the PDS; solid lines correspond to original spectra).

We test the performance of the PDS in the noisy condition. In the PDS, the CCD detector is an fundamental optical elements. Therefore, we consider the robustness of the PDS when CCD has noise.

In the noisy test, a spectral range of 400nm-700nm was selected shown in Figure 2e. The image of the hole is the same hole as shown in Figure 2a. Due to the noise existed in the CCD, the diffraction intensity patterns $I(m\Delta x, n\Delta y)$ of the light spectrum were simulated according to the following equation:

$$I(m\Delta x, n\Delta y) = \varepsilon_{rand,i} \cdot \left(10^{0.05*SNR} \Big/ \sum_{\lambda_i} a_{mn}^{\lambda_i} E(\lambda_i)\right) + \sum_{\lambda_i} a_{mn}^{\lambda_i} E(\lambda_i) \qquad (6)$$

where the first item in the right hind side of Equation(6) represents the noise part generated by the CCD, the second term in the right hind side of the Equation(6) represents the signal part generated by the spectral to be detected. $\varepsilon_{rand,i}$ are random numbers, one for each pixel, sampled from a normal distribution centered at zero, with $\sigma$ =0.1 representing different(measurement) error levels. *SNR* represents the signal to noise ratio (*SNR*) of the CCD applied in the PDS (The normal definition of *SNR* is *SNR*=20 lg (*S/N*), in which *S* represents the signal intensity and *N* represents the noise intensity). At present market, the *SNR* value of 10bit, 12 bit, 14 bit, and 16 bit CCD are normally 70db, 80db, 90db, and 96db respectively. In the reconstructions, 240 pixels are selected from the measured intensities $I(m\Delta x, n\Delta y)$ to reconstruct the spectrum. These intensity of selected pixels are substituted into Equation (1) to reconstruct the light spectrum using least-squares linear regression. At the noise level of SNR=96*db*, the reconstructed spectrum matches the original light spectrum perfectly (Fig.3a). At the noise level of SNR=90*db*, the error between the reconstructed spectrum and the original spectrum is extremely small (Fig. 3b) and the original spectral and reconstructed spectral coincided very well. As the noise level rises to SNR=85*db*, the differences increase little but the reconstructed very well(Fig.3 c). As the error level increase to SNR=80*db*, the error becomes large in some points but major peak information can still be obtained from the simulation (Fig. 3d). Further strategy for reducing the reconstruction error is to use more sophisticated reconstructions algorithms [15-17].

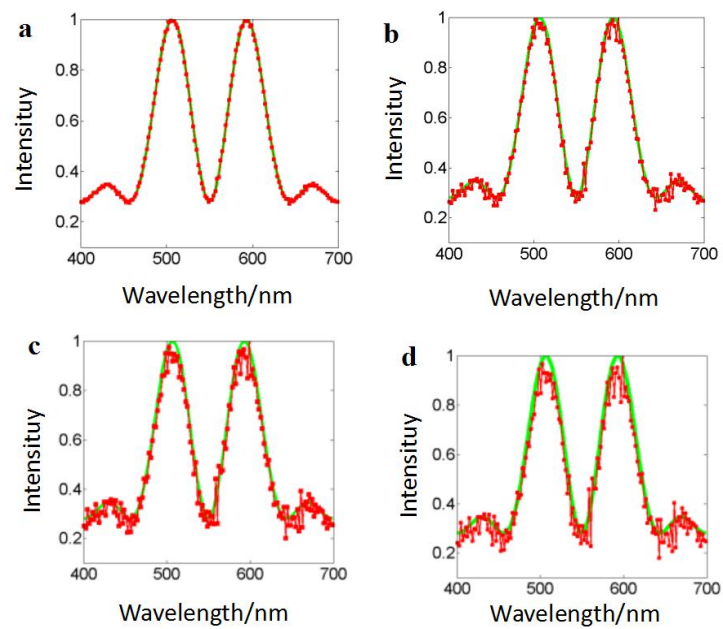

**Figure 3| Noisy experiment tests of the PDS.** The green line in a-d represent the the original light spectrum. The red lines with square in a-d represent the reconstructed light spectrum by applying the PDS when the SNR of the CCD applied is 96*db*, 90*db*, 85*db* and 80*db*, respectively.

We test the pupil diversity in the PDS. The function of the pupil in the PDS is to make the incident spectral diffract on it. Thus, the shape of the hole can be of any shape only if the spectral light diffracts on it (Note that the hole pupil is just an example, as was the given schematic in Figure 1b and Figure 2b). The pupil of the HDS has great diversity, such as rectangle hole, triangle hole, the star hole, or even the hole array, and any other images. For the sake of brevity, we only show three types of pupils in Figure 4. The size of the pupils in Figure 4 is still 1.024mm× 1.024mm. All the three pupils can obtain very well measurement results similar to those presented in Figure 2e-f, which are not shown here for brevity).

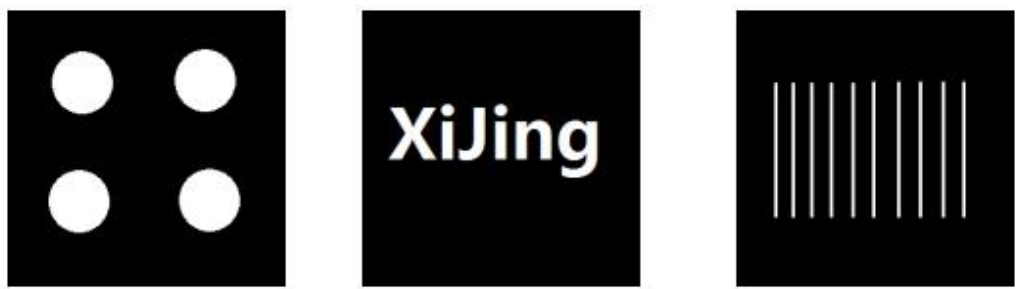

**Figure 4| Diversity of the pupil in the PDS**. a, the pupil is 2×2 holes array, b pupil is the the Chinese phonetic of XiJing. c, pupil is the grating.

## 5. Discussion

From the structure of PDS, we can see that this method has the following five outstanding advantages: **1.Low cost.** The manufacture of the hole is particularly simple, which is only need to punch a small hole on an opaque material. Since the opaque material is cheap and easy to access in the market, the cost of the PDS is approximately equal to the cost of a detector. **2. Small size and light weight.** Since the size of both the hole and the detector in the spectrometer can be made very small and thus has very light weight, the spectrometer can also be particularly small in size, and even can be integrated to the smartphone. **3. Wide Spectral range** As described above, the hole is completely light-transmissive, all the spectra (from the X-ray to the infrared ray) entering the spectrometer can reach the detector(compared with absorptive filters, their spectral range is limited because only a limited band spectrum can transmitted though it and reach the detector). Therefore, the spectral range of the spectrometer will improved to an ideal state: As long as the detector can detect the spectral range, is the method of spectral detection range. 4. **A adjustable wide angular distribution**. The PDS can analyze light from a source with a wide angular distribution while maintaining the spectral resolution. For example, we can use an integrating sphere between the light source and the spectrometer to increase the etendue of the PDS. We also can use the **diffraction optical elements** (pure phase elements) or **the multiple hole** as the pupil to increase the luminous flux in the PDS. **5. High Photon Efficiency.** The photon efficiency is close to 100%. Since the hole pupil is completely transparent, the light energy entering into the spectrometer reaches the detector completely and thus the photon efficiency can reach 100% percent.

Further, the PDS is promising high-performance micro spectrometers because spectral resolution and spectral range can be increased simultaneously, simply by increasing the number of pixels selected in the diffraction pattern. For example, if the CCD array has $10^6$ pixels (1000×1000 pixels), the spectral range can be divided into $10^6$ copies in theory. In practice, the spectral resolution can be dynamical adjustable. For example, one can select the number of pixels according to the resolution requirements. Assume that the spectral range of the spectrometer is from 100nm to 1100nm. If the spectral resolution requirement is 1nm, we can arbitrary choose $10^3$ pixels (or more pixels) from the total $10^6$ pixels of the diffraction patterns.

**6. Conclusion**

In conclusion, we propose the third principle of spectrometer. The PDS not only abandons the elaborately designed the filter and gratings, but also can ensure the spectral range and resolution simultaneously. A series of simulation results demonstrated the feasibility of the PDS principle. Owing to its low price, small size, light weight, and simple setup, it will become a promising spectrometer and an indispensable tool to enter into ordinary family to change human life [15-17].

**Acknowledgements** Tuo Li appreciates the support from the supported by grants from the Foundation of Education Department of Shaanxi Provincial Government (17JK1165).